\def\lsim{\hbox{ \rlap{\raise 0.425ex\hbox{$<$}}\lower 0.65ex\hbox{$\sim$} }}
\def\gsim{\hbox{ \rlap{\raise 0.425ex\hbox{$>$}}\lower 0.65ex\hbox{$\sim$} }}
\def\etal{et al.~}
\def\fe{{$[Fe/H]$ }}

\def\ro{$\log \rho_0$ }

\def\2p{{~$2^{nd}P$~}}
\def\p{{$(B-V)_{peak}$~} }
\def\bu{$B2-R/B+R+V$ }
\def\b{$B2/B+R+V$ }
\def\zi{$B-R/B+R+V$ }

\def\hst{{\it HST} }
\def\ie{{\it i.e.} }

\documentstyle[aj_pt4]{article}
\input epsf.sty

\accepted{by The Astronomical Journal}
\journalid{The Astronomical Journal}{}
\articleid{}{}

\vfill
\slugcomment{}

\lefthead{Buonanno et al.}
\righthead{HB morphology in Galactic GCs}

\begin{document}

\title{Horizontal Branch Morphology in Galactic Globular Clusters:\\ 
Dense Environment is ``a" Second Parameter}

\author{R. Buonanno and C. Corsi}

\author{Osservatorio Astronomico di Roma, Roma, Italy}

\author{M. Bellazzini\altaffilmark{1}, F.R. Ferraro}

\author{Osservatorio Astronomico di Bologna, 40126 Bologna, Italy}

\and 

\author{F. Fusi Pecci\altaffilmark{2}}

\author{Stazione Astronomica di Cagliari, 09012 Capoterra, Cagliari, Italy}

\altaffiltext{2}{on leave from: Osservatorio Astronomico di Bologna, 40126, 
Bologna, Italy}
\altaffiltext{1}{send comments to Michele Bellazzini, 
bellazzini@astbo3.bo.astro.it}


\begin{abstract}
The Horizontal Branch (HB) morphology in the color --
magnitude diagram of the Galactic globular clusters depends on many
factors, and it is now firmly established that the so-called 
"Second Parameter" is not just the cluster age as claimed for several years.

As a part of a wider program devoted to the search for the physical processes
driving the Horizontal Branch Morphology, 
we re-address here the problem of the extension of blue HB tails
by introducing a new quantitative observable, \b, where $B2=\{Number~ of
~HB ~stars ~with ~(B-V)_0<-0.02\}$.

We demonstrate that the environmental conditions within a cluster 
clearly affect its HB morphology, in the sense that, in general, the higher 
the cluster central density the higher is the relative
number of stars populating the most blue region of its HB.

\end{abstract}


\section{Introduction}

The so-called {\it Second Parameter} (\2p) effect in the classification 
of the Horizontal Branch (HB) morphology of Galactic globular clusters (GGC's) 
is a long-standing problem in stellar astrophysics (see  
\cite{z93,f93,ldz}, hereafter LDZ, \cite{cdf94,p95,sbv96,f96a} for recent 
reviews).
The name originates from the observation that at least one parameter other
than metallicity (the classical {\it first parameter}) has to affect the 
HB star distribution in color (and thus, temperature) in the Color-Magnitude
Diagram (CMD).

The understanding of which parameters and to what respective extent
they generate the observed HB morphologies has an
ubiquitous impact on many astrophysical topics, ranging from stellar 
evolution (see \cite{rfp} and \cite{ddro} for useful references) to galaxy 
formation and evolution (\cite{m93,fbcf,z96}).

Due to the success of the "Searle and Zinn" scenario for Galaxy
formation (\cite{sz}, hereafter SZ; LDZ, \cite{z93,z96}), the consensus
on the hypothesis that {\it the age is the only dominant \2p in the Galactic
Halo} has continuously grown in the last years. However, it is important
to stress that while this idea  seems to have been a very fruitful
base for the interpretation of a number of observational evidences in the
SZ framework, a  clear-cut direct confirmation of the hypothesis
is still lacking (see \cite{fbcf}).

Furthermore, many authors have argued that the variety of 
observed HB morphologies 
cannot be entirely (and {\it globally}, in the sense defined by \cite{l93}) 
described in terms of metallicity and age (\cite{f96a} and references
therein), and now at least one firm evidence in this sense has come out
from recent HST observations (\cite{sbv96}). 
We will present a detailed discussion of the \2p effect, within the
framework coming out from the latest results, in a dedicated
paper (\cite{f96b}, hereafter F96, in preparation). Some anticipations
can be found in Fusi Pecci et al.\ 1996a. 

In this note, in particular, we deal with the problem of 
putting into clear-cut evidence the connection between
stellar-density conditions in globulars and the presence and extension of 
{\it blue tails} (\cite{c94}) in the Horizontal Branches of their CMDs,
an idea we originally put forward 
about ten years ago (BCFP85, \cite{f87}) and that nowadays 
seems ever more worth of detailed study. 
Relevant indications about the existence of such a connection have already
been presented by our group (F93) but they {\it (a)} were based on set 
of HB morphology parameters that are hard to measure in many clusters,
and {\it (b)} were possibly subject to a spurious effect described
by van den Bergh and Morris (1994).

Based on a new parametrization, which clearly overcomes both the quoted
problems, we put on a firmer and quantitative ground the results of F93,
and we show that
the relative number of stars populating the HB blue tails
is strongly correlated to the cluster central density.

\section{Parametrizing HB morphology: the blue tails}

One of the main difficulties in studying the HB star distribution is to
provide a satisfactory, quantitative, description of its morphology
(see \cite{bia}).
As well known (\cite{rc}), the problems arise mainly 
from the fact that in the traditional (V, B-V) or (V, V-I) planes 
the blue HB tails are actually "vertical", mostly due to the increasing
bolometric correction with decreasing color index (\cite{rc1}, F93). 
Furthermore, even if the HB star distribution were to be 
Gauss function or some other unimodal function (\cite{bob73}), 
to yield a complete quantitative description one must know the location
of the peak and some kind of dispersion around it, \ie two parameters at least
(\cite{d95b}).

For instance F93 noticed that the most widely used HB morphology parameter,
\zi (\cite{z86,l89}) --where B, V, and R are the HB populations bluer, 
within, and redder than the HB instability strip, respectively-- is 
useful in ranking the various HB distributions according to 
the positions of the their peaks (see also LDZ), but it is almost insensitive 
to any variation of their extension. In particular, it is completely unable to
distinguish {\it (a)} between distributions which have similar peaks but very 
different blue tail populations and {\it (b)} between distributions that
lie only in the region of the CMD which is bluer than the instability strip
(see also \cite{d95b}).

With the aim of circumventing this problem, we (F93) proposed some "quite
crude" empirical observables ($BT$ and $L_t$ --the lengths of the 
blue tails and of the whole HB distributions, measured in arbitrary units) 
devoted to rank the observed HB's 
and their blue tails according to their color\footnotemark{ }
\footnotetext{Actually, it is not precisely the "color", as 
in the $(V, B-V)$ plane the blue HB ridge-lines are 
not $V\simeq const$ straight lines. In fact they are curves, 
with the star temperature increasing along the "red-to-blue" direction.
F93 tried to trace the covered ranges in temperature, 
by measuring the extension of the curve ridge-lines themselves.}
extension.
We believe this additional information to be of fundamental importance.
In fact, while the location of the peak of the HB distribution in
color presumably reflects the effects due to basic "average"
parameters (metallicity, age, mean mass loss, etc.) common to all cluster
members, the spread around it
yields a {\it direct} measure of the spread in total mass loss and/or
core mass (eventually due to unknow mechanisms, like for instance spread 
in core rotation, special mixing, tidal stripping, etc.).
It seems very likely that the two items are strictly 
connected, and any quantitative observational study of the color spread and 
of its possible correlation with any cluster parameter could set valuable 
constraints on the whole "mass loss affair" and, in turn, on the \2p--problem.

\subsection{Definition of a new index}

Using the quoted parameters, F93 found that the color extension of the
HB distribution is correlated with cluster central density 
($\rho_0, L_V/pc^3$), 
in the sense that more centrally concentrated clusters tend to 
have bluer HB-morphologies and often present extended blue HB tails
(see also \cite{c94}).
van den Bergh and Morris (1994, hereafter vdBM) have questioned this result, 
suggesting 
that "the apparent correlations between the length of the HB and the cluster
luminosity and density may be affected by observational bias". In particular,
because of the clear correlation existing between luminosity and central 
density (\cite{dm94}), in their view the correlation found 
by F93 could be a spurious effect reflecting the fact that "one can trace
HB's over greater lengths in rich clusters than in poor ones".

To examine the criticism raised by vdBM and to obtain
quantitative information on the blue tails not affected by the quoted 
possible bias (hereafter {\it vdBM-bias}), we define the index \b as the 
ratio of the number $B2$ of observed HB stars bluer than $(B-V)_0=-0.02$,
to the number of all the stars belonging to the HB distribution,
$B+V+R$. Further operative details can be found in the preliminary reports 
by Buonanno (1993), Buonanno and Iannicola (1994) and Fusi Pecci et al. (1996a).

Being normalized to a quantity directely related to the total luminosity of 
the sampled population (\cite{rfp}), \b is necessarily free from the quoted 
{\it vdBM-bias}. Nevertheless, for sake of checking and recalling also
the caveats put forward by Ferraro, Fusi Pecci and Bellazzini (1995), 
a wide set of
simulations have been performed to test the variation of 
\b with the cluster integrated luminosity. These simulations
show that, for any fixed ``realistic'' couple of the peak
and of the standard deviation of the distributions (assumed to be Gauss
functions), \b turns out to be 
insensitive to any change of the cluster total luminosity within the 
range actually spanned by Galactic globulars (\ie, $-4\le M_V\le -10$; 
more details about the simulations will be given in the forthcoming paper
F96). 

Beside \b defined above, we will use below also another parameter, \p, 
defined and measured by F93 to determine the actual  location 
(in $(B-V)_0$) of the peaks 
of the HB stellar distribution. F93 already showed that this observable
strongly correlates with \zi, and substantially yields similar basic
information.

\placetable{tbl1}

\section{Database and results.}

\subsection{The HB data}

Our complete database collects 63 halo globular clusters for which the
new HB parameters can be measured on the basis of available
photometric studies. 
The entries are limited to clusters with $[Fe/H]<-1.0$ because 
the observed clusters with higher metallicity all have
very red, stubby HB's. This is due to the very non-linear response to mass 
loss by stars 
in different metallicity regimes (see F93, Figure 1,2 and DDRO, Figure 6 
and the discussion therein). In fact, according to the current 
standard and canonical HB models, a star with such a high metal content
need to lose as much as $\sim 0.3$M$_{\odot}$ to be located in a 
ZAHB position bluer than the instability strip. This same phaenomenon
strongly limits the capability of a metal rich population to produce blue
HB stars (but see sect. 4). Since our index is devoted to study blue
morphologies and it is "blind" to very red ones, the inclusion of the
metal rich cluster in the sample would have been nonsense.

The database is presented in Table 1, whose first column reports
the identification number or name of the clusters. 
Metallicities (\fe, $2^{nd}$ col.) and central densities (\ro, $3^{rd}$ col.) 
are taken from Djorgovski (1994).
The de-reddened colors of the peak of the 
HB-distributions (\p, $6^{th}$ col.) are drawn from F93 (44 entries).
The new HB-morphology index \b, listed in the $4^{th}$ column for all 
considered clusters, has been obtained by star counts on the CMDs referenced 
in the $7^{th}$ column.
As an independent check we calculated also the parameter \zi, and found 
excellent agreement with the counts of LDZ.

A fully reliable extimation of the uncertainties affecting indexes 
as \zi or \b is an almost unaffordable task. In fact this kind of
indexes suffer the effects of three main indeterminacy souces: 
(i) the Poisson noise
related to star count statistics, and in turn to the number of sampled
stars; (ii) the precision of the photometry; note that this item is particulary
difficult to deal with when a large set of measures are collected from different
photometric studies, each with a different photometric precision;
(iii) completeness problems. At present, the only attempt to extimate
\zi errors has been made by LDZ which, based on synthetic HB calculation,
find only "...estimates of the uncertainty due to the random distribution of
stars along the evolutionary tracks" and assume that this figure is 
nearly half of the total uncertainty affecting the measures of the index.
Since the error source described at item (i) above, is the only one
which can be somehow controlled from an empirical point of view, we
prefer to list the quantity $\sqrt{B+R+V}\over{B+R+V}$ (Tab. 1, $5^{th}$ col.)
which ranks all \b entries according to the total number of sampled
HB stars.
These figures  are just indicative of the quality of
the population sampling, and are not intended to describe also the 
photometric quality and/or completeness of the sample. 

Actually, the determination of precise errors is not so crucial for the 
present purposes as we are not 
aiming at deriving any physical quantity from our indexes. We rather
use them as a statistical data-set against which to test the likelyhood
of a particular hypothesis, i.e. to estimate
the probability that a given correlation and/or association can be 
originated just by chance.

\subsection{Dense environment and blue tails}

In order to search for a clear-cut evidence of the link possibly connecting
intrinsic structural parameters of the considered clusters to
existence and extension of the blue HB tail,
we divided the available sample into two groups:

$\bullet$ ~~HD-group: {\it high density} clusters, with 
$log \rho_0>3$ (37 objects)

$\bullet$ ~~LD-group: {\it low density} clusters, with 
$log \rho_0\le 3$ (26 objects).

We performed a Kolmogorov-Smirnov test to verify whether the two 
groups display distributions in \b which are compatible with drawing 
them from the same parent population. 
{\it The probability that the two sub-groups are extracted from the 
same parent population in \b} turns out to be $0.02 \%$.
Figure 1 (left panel) shows that the distribution of the LD-group 
is significantly skewed towards low values of \b. In fact, $85\%$ 
of the sample have $B2/B+R+V\le 0.2$.
At the same time, it is particular worth noting that {\it the two 
sub-groups of clusters have indistinguishable metallicity distributions}
(Figure 1, right panel). This ensure that the above result is free from 
any spurious effect due to sampling in  metallicity.

\begin{figure}
\epsscale{.8}
\plotone{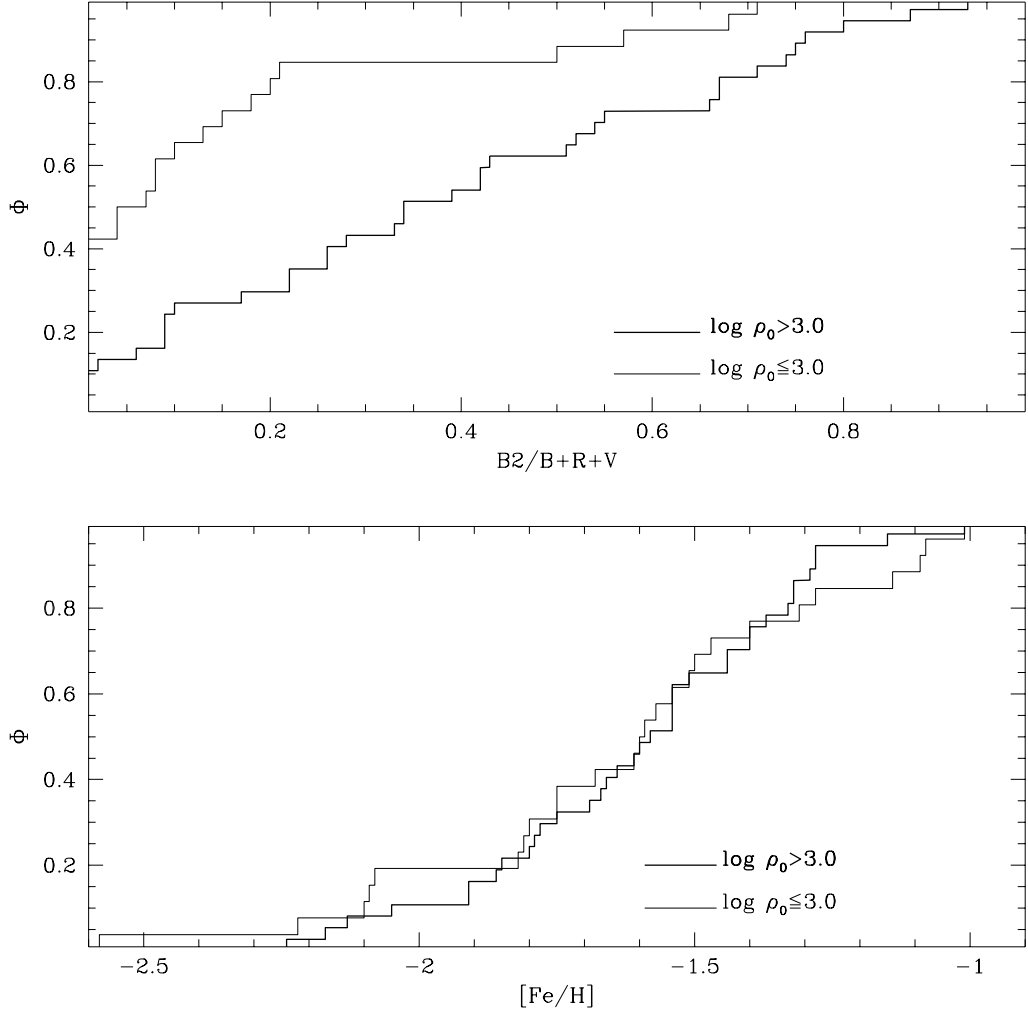}
\caption{
Upper panel: the cumulative distribution of the $B2/B+R+V$
index for the High Density group (thick line) compared
to that for Low Density group (thin line), see Sect. 3.1. 
The HD-group distribution is significantly skewed towards higher 
values of the \b index with respect to the LD one.
Lower panel: (same graphic labels) the cumulative 
metallicity distribution of the two groups are undistinguishable.
\label{fig. 1}}
\end{figure}

As a first conclusion, we can thus say that Figure 1 shows {\it direct}
evidence (in the sense that it is fully "model-independent") that
{\it clusters having the same metallicity, but different intrinsic
concentrations behave differently as far as "blue" HB morphology is 
concerned} and, more specifically, {\it stars harboured in very dense 
cluster environments have higher probability to populate
the most blue side of the Horizontal Branch distribution in a CMD}.

Our analysis can now be pushed a bit further. We have already noted that
the index \b presumably depends  not only on the extension of the blue
side of the HB distribution but could also be somehow influenced 
by the position of the peak. The main reason why this is expected to occur 
is very simple: given two unimodal distributions with the same
color spread, the number of stars falling blueward of the $(B-V)_0=-0.02$
will be higher for the distribution that display the "bluer" peak.
Hence, it is worth trying to disentangle the two dependences, making use 
also of the \p parameter. By doing this, we simultaneously test 
our suggestion that the "dense environment effect" mostly acts 
on the spread of the distribution, rather than on the HB peak.
If we assume that the peak reflects the {\it mean} properties 
of the HB distribution, it seems very likely that its position is driven 
mainly by the quoted average parameters which are shared by the whole 
considered population (i.e. metallicity, age etc., as said). So, taking
into account the influence of \p on the measured \b we can hope
to trace better the {\it real} color spread.

As a preliminary result note that, as already pointed out by F93,
\p do not correlate with \ro, while there is a correlation between
\p and \fe (see FP93, Figure 8). Thus, at a zeroth approximation,
\p appears to trace the {\it first parameter}, \fe.

In Figure 2 we have plotted \b {\it vs.}
\p. Inspecting the plot, one immediately sees two regimes. 
The \b index is quite insensitive
to variations in \p if \p is larger than $\sim 0.3$mag. This is so
because the the peak of the distributions are so red that the \b
index is unable to provide any significant ranking. 
For \p$<0.3$ mag, the correlation between \bu and \p becomes evident
and apparently linear. We performed thus a linear fit to the data 
within this linear range and calculated the residuals
$[(B2/B+R+V)_{obs}-(B2/B+R+V)_{fit}]$. If our claim that the
population of the blue tails is driven by a factor other than
metallicity and age is correct, we must find that these residuals
correlate with some other physical quantity. 

\begin{figure}
\epsscale{.8}
\plotone{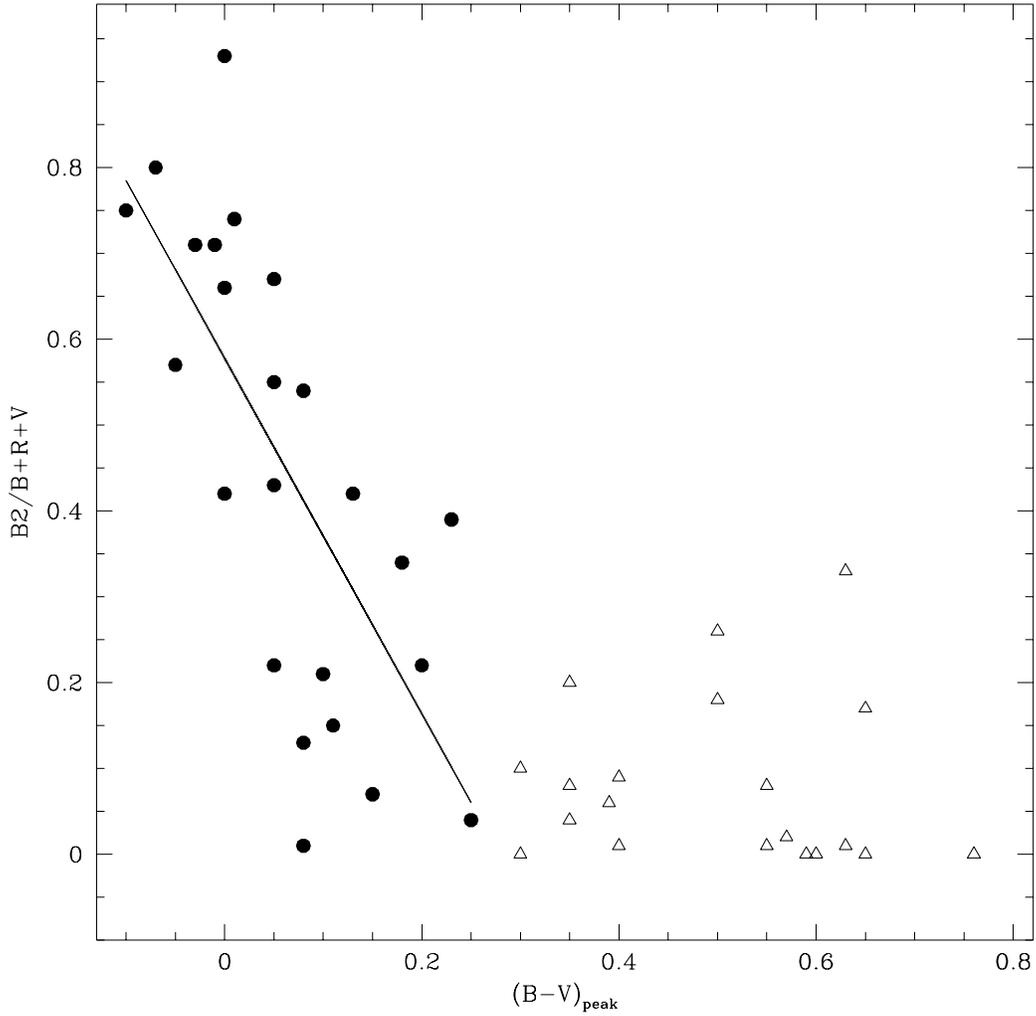}
\caption{
The index $B2/B+R+V$ versus the color of the peak of the
Horizontal Branch distribution, \p . Filled dots identify the clusters
having a peak position sufficiently blue to make the "amplification
effect" due to dense environment highly evident and yield
a significant correlation between \p and \b. The line is just a 
linear regression fit to the data. Open triangles represent
the clusters that have been excluded from the linear fit because
of their red peak (\p $>0.3$)
\label{fig. 2}}
\end{figure}

\begin{figure}
\epsscale{.8}
\plotone{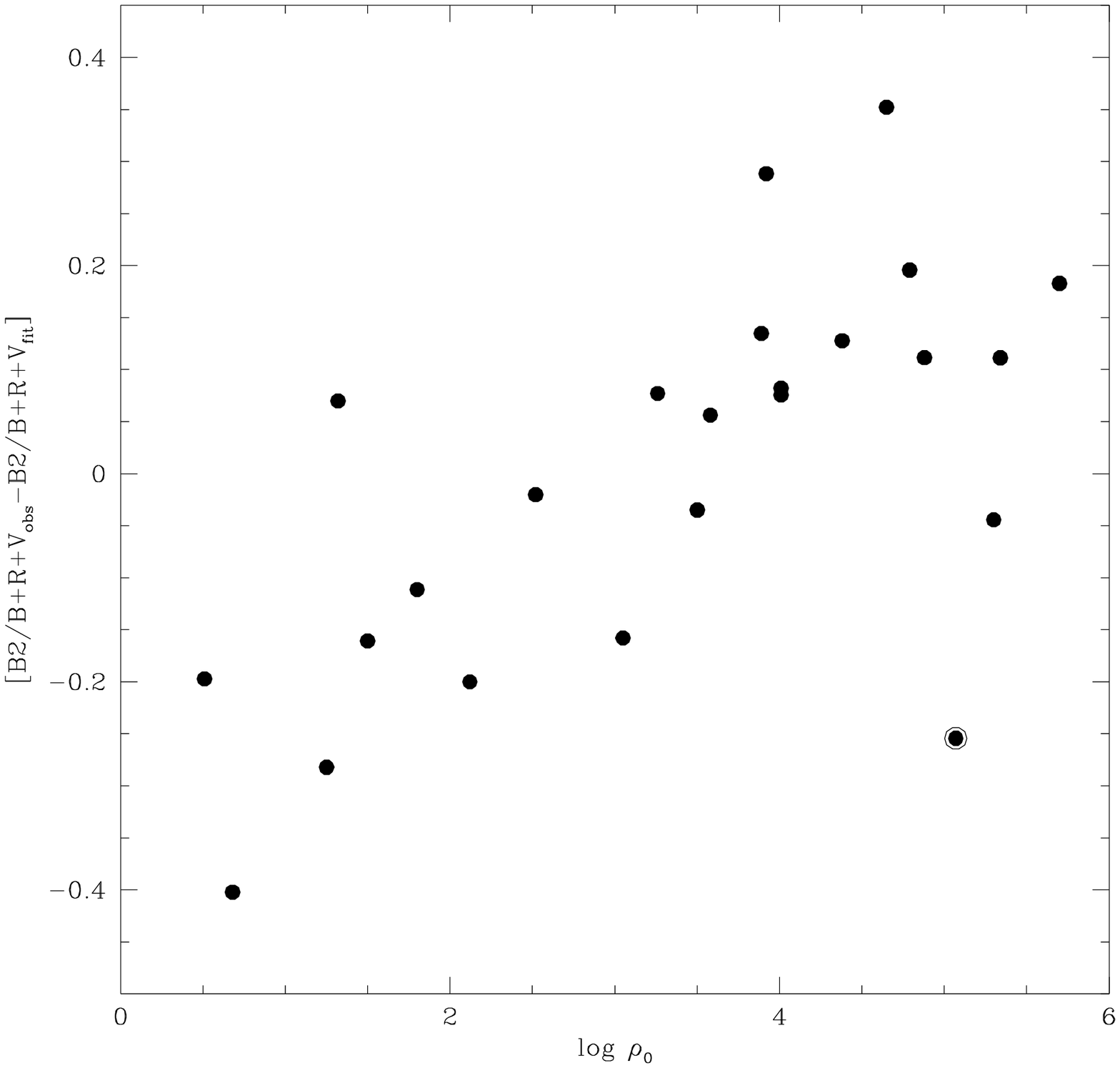}
\caption{
The residuals of the fit shown in fig. 2 are found to
correlate with \ro. The Spearman's rank correlation coefficient is 
$s=0.61$ and the probability that the two involved quantities are 
{\it uncorrelated} is less than $1 \%$. The outcircled point corresponds
to NGC 7099.
\label{fig. 3}}
\end{figure}

In Figure 3, it is shown that the residuals display a
clear-cut correlation  with \ro. The Spearman's rank correlation coefficient
is $s = 0.61$, and the probability that the involved parameters
are actually uncorrelated is less than $1\%$.
The observed dispersion
of the relation can be completely accounted for by two different factors:
\begin{itemize}
\item[a)] the observational errors which can well be, as said, 
larger than the figure quoted in column 7 of table 1 due to errors and 
insufficient completeness of the HB star photometry. Note that,
while it seems unlikely that photometric errors do affect significantly
the actual stellar counts, completeness can play a significant r\^ole 
when dealing with stars that can be as dim as the Turn Off of the 
considered cluster (see for instance the case of NGC 6752, 
Buonanno \etal 1986).

\item[b)] the quoted non-linear modulation of the response to mass loss settled
by the metallicity, i.e. populations of different metallicity are expected 
to display quite different color-shift response to the same amount of 
mass loss (see F93 and DDRO). This effect must contribute in smearing the
observed correlation, given the wide metallicity range ($-1.0<[Fe/H]<-2.5$)
covered by the considered sample.
\end{itemize}

\noindent
The only point whose deviation from the mean overall trend may
be considered significant (the outcircled dot in fig. 3), 
represents the very metal poor cluster NGC7099 ($[Fe/H]=-2.3$). 
At present, we can just guess that its \b measure has been probably
underestimated because of incompleteness or of some unknown systematic 
error in the photometry. 

From the above results one can thus draw the second conclusion:
Unless cluster stellar density depends on age via a very special (unknown)
connection, {\it some physical process which affects the evolution 
of stars living in a very dense environment must be at work within 
the cluster population}. 
Furthermore, this mechanism shows its effects mostly on the shape of 
the HB-distribution, enhancing the relative number of stars in the blue 
HB side, wherever the peak of the distribution itself is located
(but, of course, with an "amplification factor" depending on the metallicity
regime, see F93).

\section{Discussion and future prospects}

In the latest years, the understanding of the very blue Horizontal 
Branch stars has been growing in importance due to the impact it may
have on topics like the quoted long-standing \2p--problem, the
connections between stellar evolution and dynamics 
(see \cite{bai}), and the possible 
explanation of the UV-excess detected in the integrated light of many 
ellipticals and of spiral galaxy nuclei (see for references 
\cite{gr}, Dorman et al. 1993, 1995, \cite{bcb96}). 
In these section we discuss briefly some of the spin-offs of our
results and add few speculations related to some of the above-quoted topics.

\begin{itemize} 

\item[1] The use of the HB morphology as a plain age indicator is 
no more justified, at least if only the {\it standard} parametrization
(i.e. the \zi index) is adopted. This is particulary true when dealing
with ``blue morphologies'': in this cases the effects of old age cannot
be distinguished from the effects of high density conditions. We demonstrated
that a more suitable parametrization is clearly needed to disentangle the
various effects.

\item[2] The behaviours of the peak of the HB distribution and of the 
{\it blue tail} are apparently decoupled (see fig.2, 3).
This strongly suggests that the mechanism connecting {\it density} and 
{\it blue tails} is a rather random one, which seems   
{\it (a)} to be effective in perturbing only a fraction of the cluster 
stars, and {\it (b)} to act with different effectiveness on each
of the perturbed star. Concerning the processes for the production of
extremely blue HB stars, either the mechanisms reviewed by
Bailyn (1995) (based on binary interactions) or the one proposed by 
BCFP85 seem consistent with the above-described scenario.

\item[3] Assuming for instance that the process linking the production of
blue HB objects and stellar density grows up in its efficiency in a 
continuous way with increasing density, one could also imagine that
in some metal rich bulge clusters, particulary extreme density
conditions (as, for example the occurence of a gravothermal collapse 
phase) may yield the amount of mass loss required to induce the 
{\it blue transition} of the whole HB, or of parts of it. 
If such clusters do exist, their HB has presumably
a bimodal (or multi-modal, see DDRO) overall morphology, 
with the expected red clump populated
by the stars "not-" or "less-affected" by this (unknown) mechanism
which enhances mass loss, and a long blue tail (probably separated by a 
gap, see DDRO, Figure 7) populated by the "perturbed" stars. 
Consequently, it is also possible that some metal rich clusters already studied
do display a blue HB population which has escaped detection
because it is very faint in V, perhaps fainter than the Main Sequence
Turn Off. The search for such features can surely be accomplished 
with the \hst facilities, and the preliminary results presented
by Piotto \etal (1996b) seem to be a first direct confirmation of
the validity of this hypothesis. 

\end{itemize}

\acknowledgments

We are grateful to Ben Dorman for a critical reading of the early draft,
and to Sidney van den Bergh for useful discussions.
The financial support of the {\it Ministero della Universit\`a e della
Ricerca Scientifica e Tecnologica} is gratefully acknowledged.


\clearpage







\begin{deluxetable}{lcccccc}
\tablenum{1}
\tablewidth{0pt}
\tablecaption{Data for Galactic Globular Clusters}
\tablehead{
\colhead{Name}                       & \colhead{[Fe/H]}             &
\colhead{$\log \rho_0$}              & \colhead{$B2\over{B+R+V}$}   &
\colhead{$\sqrt{B+R+V}\over{B+R+V}$} & \colhead{$(B-V)_{peak}$}     &
\colhead{Ref.}
}
\startdata
NGC 288 & -1.40&  1.80& 0.57& 0.10& -0.05& 1 \nl
NGC 362 & -1.28&  4.75& 0.01& 0.10&  0.55& 2 \nl
NGC 1261& -1.31&  2.98& 0.01& 0.06&  0.63& 3 \nl
NGC 1851& -1.29&  5.16& 0.17& 0.10&  0.65& 4 \nl
NGC 1904& -1.69&  4.01& 0.55& 0.08&  0.05& 5 \nl
NGC 2419& -2.10&  1.50& 0.21& 0.09&  0.10& 6 \nl 
NGC 2808& -1.37&  4.63& 0.33& 0.04&  0.63& 7 \nl
NGC 3201& -1.61&  2.63& 0.20& 0.08&  0.35& 8 \nl
NGC 4147& -1.80&  3.58& 0.22& 0.13&  0.20& 9 \nl
NGC 4372& -2.08&  2.18& 0.68& 0.09& --& 10 \nl
NGC 4590& -2.09&  2.52& 0.04& 0.10&  0.25& 11 \nl
NGC 4833& -1.86&  3.05& 0.42& 0.10&  0.00& 12 \nl
NGC 5053& -2.58&  0.51& 0.07& 0.15&  0.15& 13 \nl
NGC 5272& -1.66&  3.54& 0.10& 0.07&  0.30& 14 \nl
NGC 5286& -1.79&  4.19& 0.51& 0.08& --& 10 \nl
NGC 5466& -2.22&  0.68& 0.01& 0.11&  0.08& 15 \nl
NGC 5694& -1.91&  4.01& 0.66& 0.17&  0.00& 16 \nl
NGC 5824& -1.85&  4.65& 0.26& 0.08& --& 10 \nl
NGC 5897& -1.68&  1.32& 0.71& 0.13& -0.03& 17 \nl
NGC 5904& -1.40&  3.92& 0.39& 0.08&  0.23& 18 \nl
NGC 5986& -1.67&  3.24& 0.52& 0.13& --& 19 \nl
NGC 6093& -1.64&  4.79& 0.67& 0.15&  0.05& 20 \nl
NGC 6101& -1.81&  1.57& 0.10& 0.11& --& 21 \nl
NGC 6121& -1.33&  3.91& 0.01& 0.08&  0.40& 22 \nl
NGC 6218& -1.61&  3.26& 0.80& 0.09& -0.07& 10 \nl
NGC 6229& -1.54&  3.46& 0.26& 0.11&  0.50& 23 \nl
NGC 6254& -1.60&  3.50& 0.75& 0.13& -0.10& 19 \nl
NGC 6266& -1.28&  5.34& 0.42& 0.06&  0.13& 10 \nl
NGC 6284& -1.40&  4.65& 0.93& 0.09&  0.00& 20 \nl
NGC 6287& -2.05&  3.99& 0.00& 0.12& --& 24 \nl
NGC 6341& -2.24&  4.38& 0.54& 0.09&  0.08& 15 \nl
NGC 6333& -1.78&  3.58& 0.09& 0.10& --& 25 \nl
NGC 6362& -1.08&  2.23& 0.08& 0.11&  0.55& 26 \nl
NGC 6397& -1.91&  5.70& 0.74& 0.08&  0.01& 27 \nl
NGC 6522& -1.44&  5.53& 0.34& 0.09& --& 28 \nl
NGC 6535& -1.75&  2.25& 0.50& 0.25& --& 29 \nl
NGC 6584& -1.54&  3.19& 0.01& 0.08& --& 30 \nl
NGC 6626& -1.44&  4.71& 0.67& 0.13& --& 31 \nl
NGC 6638& -1.15&  4.02& 0.09& 0.21&  0.40& 32 \nl
NGC 6656& -1.75&  3.67& 0.28& 0.10& --& 33 \nl
NGC 6681& -1.51&  5.56& 0.87& 0.10& --& 10 \nl
NGC 6712& -1.01&  3.09& 0.02& 0.13&  0.57& 34 \nl
NGC 6717& -1.32&  4.60& 0.76& 0.24& --& 10 \nl
NGC 6723& -1.09&  2.71& 0.18& 0.10&  0.50& 35 \nl
NGC 6752& -1.54&  4.88& 0.71& 0.07& -0.01& 36 \nl
NGC 6809& -1.82&  2.12& 0.15& 0.07&  0.11& 37 \nl
NGC 6864& -1.32&  4.53& 0.06& 0.09&  0.39& 38 \nl
NGC 6934& -1.54&  3.43& 0.09& 0.08& --& 10 \nl
NGC 6981& -1.54&  2.26& 0.08& 0.09&  0.35& 10 \nl
NGC 7006& -1.59&  2.42& 0.04& 0.11&  0.35& 18 \nl
NGC 7078& -2.17&  5.30& 0.43& 0.07&  0.05& 15 \nl
NGC 7089& -1.58&  3.89& 0.34& 0.10&  0.18& 38 \nl
NGC 7099& -2.13&  5.07& 0.22& 0.11&  0.05& 20 \nl
NGC 7492& -1.51&  1.25& 0.13& 0.18&  0.08& 39 \nl
AM   1  & -1.01&  0.32& 0.00& 0.19&  0.60& 40 \nl
Pal  3  & -1.57&  0.04& 0.00& 0.24& --& 41 \nl
Pal  4  & -1.28& -0.24& 0.00& 0.22&  0.59& 42 \nl
Pal  5  & -1.47& -0.77& 0.00& 0.22&  0.30& 43 \nl
Pal 12  & -1.14&  0.68& 0.00& 0.38&  0.76& 44 \nl
Pal 14  & -1.60& -1.17& 0.00& 0.25&  0.65& 45 \nl
Rup  106& -1.80&  1.22& 0.00& 0.15& --& 46 \nl
Arp  2  & -1.75& -0.35& 0.00& 0.27& --& 47 \nl
IC  4499& -1.50&  1.50& 0.01& 0.08& --& 48 \nl
\tablerefs{
(1) Buonanno et al. 1984; (2) Harris 1982; (3) Ferraro et al. 1993; (4)
Stetson 1981; (5) Ferraro et al. 1992; (6) Racine \& Harris 1975; 
(7) Ferraro et al. 1990;
(8) Lee 1977b; (9) Sandage 1955; (10) Brocato et al. 1996; (11) Walker 1994;
(12) Menzies 1972; (13) Sarajedini 1995; (14) Buonanno et al. 1994;
(15) Buonanno et al. 1985; (16) Ortolani \& Gratton 1990; 
(17) Ferraro, Fusi Pecci \& Buonanno 1992; (18)
Buonanno, Corsi \& Fusi Pecci 1991; (19) Harris, Racine \& de Roux 1976; 
(20) Piotto 1996a; (21) Sarajedini 1991;
(22) Lee 1977a; (23) Iannicola 1996;
(24) Stetson \& West 1994; 
(25) Janes \& Heasley 1991;
(26) Alcaino 1976; (27) Alcaino et al. 1987; (28) Barbuy, Ortolani \&
Bica 1994; 
(29) Liller 1980;
(30) Sarajedini \& Forrester 1995; (31) Alcaino ; (32) Alcaino \& Liller 1983; 
(33) Alcaino 1983;
(34) Cudworth 1988; (35) Menzies 1974; 
(36) Buonanno et al. 1986; (37) Lee 1977c; (38) Harris 1975;
(39) Buonanno et al. 1987; (40) Aaronson, Schommer \& 
Olzewski 1984; (41) Gratton \& Ortolani 1984;
(42) Christian \& Heasley 1986; (43) Sandage \& Hartwick 1977; (44) 
Gratton \& Ortolani 1989; (45) Harris \& van den Bergh 1984; 
(46) Buonanno et al. 1990;
(47) Buonanno et al. 1994;
(48) Ferraro et al. 1995.
}
\enddata
\end{deluxetable}
\end{document}